\documentclass[conference]{IEEEtran}
\usepackage{amssymb,amsfonts,amsmath,amsthm}

\newcommand{\E}[1]{\textrm{E}\left[ #1\right]}
\newcommand{\av}[1]{\langle #1 \rangle}
\newtheorem{prop}{Proposition}
\newcommand{\etal}{\emph{et~al.}}
\def\versubmission{1}
\def\vertechrep{2}

\def\ver{1}

\newcommand{\switch}[2]{
	\ifx\ver\vertechrep 
		{#2}
	\fi
	\ifx\ver\versubmission
		{#1}
	\fi
}

\ifCLASSINFOpdf
  \usepackage[pdftex]{graphicx}
\else
  \usepackage[dvips]{graphicx}
\fi

\usepackage[font=footnotesize, caption=false]{subfig}

\begin{document}

\title{How to Network in Online Social Networks}

\author{
\IEEEauthorblockN{Giovanni Neglia}
\IEEEauthorblockA{Inria Sophia-Antipolis M\'editerran\'ee\\
Email: giovanni.neglia@inria.fr}
\and
\IEEEauthorblockN{Xiuhui Ye}
\IEEEauthorblockA{Politecnico di Torino\\
Email: yexiuhui@gmail.com}
\and
\IEEEauthorblockN{Maksym Gabielkov, Arnaud Legout}
\IEEEauthorblockA{Inria Sophia-Antipolis M\'editerran\'ee\\
Email: \{firstname.lastname\}@inria.fr}
}

\maketitle

\begin{abstract}
In this paper, we consider how to maximize users' influence in Online Social Networks (OSNs) by exploiting social relationships only.
Our first contribution is to extend to OSNs the model of Kempe \etal{}~\cite{Kempe03} on the propagation of information in a social network and to show that a greedy algorithm is a good approximation of the optimal algorithm that is NP-hard.
However, the greedy algorithm requires global knowledge, which is hardly practical.
Our second contribution is to show on simulations on the full Twitter social graph that simple and practical strategies perform close to the greedy algorithm.
\end{abstract}

\IEEEpeerreviewmaketitle

\section{Introduction}
\label{s:introduction}
The first motivation of any social network is to foster information propagation using social relationships among users.
Therefore, it is important to understand how a user can reach a large population, that is how to best exploit users' influence. 

Domingos \etal{}~\cite{Domingos01} introduced first the general problem to select individuals to spread information taking into account their influence.
Then Kempe \etal{}~\cite{Kempe03,Kempe05} defined a general optimization frameworkto solve this problem.
However, the main implicit assumption made by Kempe \etal{} is that once the influential individuals have been identified, they can be \emph{recruited} in order to spread the information of interest.
For instance, recruitment can be made with a monetary transaction.
Kempe \etal{} make two main assumptions for the recruitment process: the recruitment budget is limited, and when users are recruited, they cannot refuse (so the recruitment decision is the one of the user who wants to spread information).
However, the recruitment process in Online Social Networks (OSNs) is much different, because it is made using friend requests.
Such a recruitment has three main specificities that are not covered by the model of Kempe \etal:
i) there is no guarantee that the friend request will be accepted;
ii) a friend request is cheap and does not require monetary transaction, so dynamic strategies are possible;
iii) the metric to quantify influence in OSNs, unlike the one used by Kempe \etal, is not only the number of users that relay the information, but also the number of users that receive the information.
Cha \etal{} \cite{Cha10} give a more general discussion about possible metrics of influence in Twitter. 

In this paper, we make the following contributions.

i) We extend the model of Kempe \etal{} with the three specificities of OSNs and show that the initial result of Kempe \etal{} still holds, that is the greedy algorithm is a $(1-1/e)$ approximation of the optimal algorithm that is NP-hard.

ii) Using the complete social graph of Twitter crawled in July 2012~\cite{Gabielkov12} and consisting of 505 million nodes and 23 billion edges, we show that, if only the degree and reciprocation probability of each node~$i$ (respectively $d_i$ and $r_i$) are known and retweet probabilities are homogeneous, the simple strategy to select the nodes with the largest product $r_i d_i$ performs at most $2.5\%$ worse than the greedy algorithm.
Moreover, selecting users at random achieves similar performance when the replication probability of the cascade process is as large as~$1\%$ and only requires to know the users IDs.
Some similar results were observed by Chen \etal{}~\cite{Chen09} and Habiba \etal{}~\cite{Habiba11}, but on much smaller graphs.
 
\section{Problem formulation and analysis}
\label{s:analysis}
In this section, we start by summarizing the contribution of Kempe \etal{} on which we are building~\cite{Kempe03,Kempe05}, then we present our extension of the Kempe's model. 

\subsection{Kempe's Model}
Kempe \etal{} modeled the propagation of information in a social network using two different discrete-time models.
The first one is the \emph{order independent cascade model}.
Nodes with or without the information are respectively called \emph{active} or \emph{inactive}.
When one node $u$ becomes active at time $t$, it has one chance to influence (infect) all its non-active neighbors, who may then become active at time $t+1$.
From time $t+1$ on, node $u$ is still active but no more \emph{contagious}.
The contagion attempts from new active nodes at $t+1$ are arbitrarily ordered.
The probability of success needs to be specified in order to completely describe the model.
A quite general case is when $u$'s success probability to infect $v$ depends on the set $S$ of $v$'s neighbors that already attempted to influence $v$.
We denote such probability $p_v(u,S)$.
In a \emph{decreasing cascade model} this probability is non-increasing in $S$, that is $p_v(u,S) \ge p_v(u,T)$ whenever $S\subseteq T$.
This corresponds to the fact that the more nodes have already tried in vain to infect $v$, the less likely $v$ is to be influenced by other attempts.
Starting from an initial set of active nodes $A$, the process will stop in at most $n-1$ steps.
The main performance metric of interest is the final set $\phi(A)$ of active nodes or better its expected size $\E{|\phi(A)|}\triangleq \sigma(A)$.
The second model is the \emph{general threshold model}.
In this case, each node has a monotone activation function $f_v: 2^V \to [0,1]$ and a threshold $\theta_v$ chosen independently and uniformly at random
from the interval $(0,1]$.
A node $v$ becomes active at time $t+1$ if $f_v(S)\ge \theta_v$, where $S$ is the set of active nodes at time $t$.
Interestingly, Kempe \etal{} \cite{Kempe05} show that the two models are equivalent, in the sense that for any activation functions $f_v(.)$, there exist corresponding activation success probabilities $p_v(.)$ such that the probability distribution over the final active sets $\varphi(A)$ is the same under both models.

The optimization problem introduced by Kempe \etal{}~\cite{Kempe03} is to choose the initial set $A$ under the constraint that $|A|\le K$ so that the expected size of the active nodes' final set is maximized.
They show that the problem is NP-hard, but that a natural greedy heuristic reaches a $(1-1/e)$ approximation factor for the decreasing cascade model (and for the corresponding general threshold model).
The greedy heuristic simply incrementally increases the set $A$ starting from an empty set and adding at each time the node $v_i$ that maximizes the marginal gain $\sigma(A \cup \{v\})-\sigma(A)$.
If at each step the selected node is a $1-\epsilon$ approximation of the best node, then the greedy algorithm achieves a $(1-1/e-\epsilon')$ approximation factor, where $\epsilon'$ depends on $\epsilon$ polynomially.
The key for proving this result is to show that $\sigma(A)$ is a non-negative, monotone, submodular function on sets\footnote{ A set function $f(.)$ is submodular if $f(S \cup \{z\})-f(S)\ge f(T \cup \{z\}-f(T)$ whenever $S \subseteq T$ and it is monotone if $f(S \cup \{z\}) \ge f(S)$ for each $S$ and $z$.}, then the conclusion about greedy algorithm's approximation ratio follows from known results on such functions \cite{Nemhauser1978,Nemhauser1988}.

\subsection{Extension of Kempe's Model to OSNs}
In the following, we extend the model of Kempe \etal{} to the specificities of OSNs, and, for the sake of simplicity, we refer to Twitter in our description of the problems.
Twitter is one of the largest social networks, but it differs from other social networks, such as Facebook, because it uses exclusively directed edges (arcs) among accounts.
Twitter has no notion of bidirectional friendship, but it allows users to \emph{follow} other users, i.e.,~to subscribe for their messages.
Following does not require any approval from the user being followed.
If Alice follows Bob, then Alice is called a \emph{follower} of Bob and Bob a \emph{following} of Alice.
Twitter users can \emph{retweet} received tweets, that is forwarding the
tweets to their followers.
In this paper, we use the notation $(V, E)$ to refer to the Twitter social graph, where $V$ is the set of Twitter users and $E$ is the set of directed edges.
We orient the arcs in such a way that they show the tweet propagation direction, e.g., if A follows B, the arc is directed from B to A, because A receives tweets from B.

Kempe \etal{} for their influence maximization problem made the implicit assumption that once the influential individuals have been identified, they can be \emph{recruited} in order to spread the information of interest.
Recruitment is costly (in terms of money or social investment), so the available budget limits the number of individuals to be selected to~$K$. 

The recruitment process in OSNs is different from the one described by Kempe.
Indeed, a user $v$ is recruited by $u$ when $v$ receives information from $u$, i.e., in Twitter terminology, $v$ follows $u$.
This recruitment process in OSNs leads to three specificities.
First, even if the user follows the most influential individuals, there is no guarantee that they will follow him back (that is, be recruited), we call this specificity the follow back problem.
Second, following is a quite cheap operation in OSNs so that more aggressive dynamic strategies are feasible, we call this specificity the dynamic problem.
Third, we quantify the influence not only by the number of individuals who actively replicate the information but also who can see the information because they follow the original tweeter or one of the retweeters, we call this specificity the reader problem.
In the following, we extend the model of Kempe \etal{} to tackle these three specificities. 

\paragraph{\textbf{Follow back problem}}
We define the influence of a user $u$ as the expected number of users who retweet $u$'s tweets.
We stress that, in our model, a user can be retweeted only if some other users follow him and decide to retweet his tweets.
In the following, we show how we can extend Kempe's model to this scenario.

Let a node be active if it has read the tweet and decided to retweet it.
Then the original influence maximization problem can be rephrased as follows: how to choose $K$ nodes that should initially tweet the message in order to maximize the expected number of retweets.
In this case, $p_v(u,S)$ is the probability that node $v$ reads and decides to retweet the message tweeted or retweeted from $u$, given that the nodes in $S$ have already tweeted or retweeted it. 

In the original problem formulated by Kempe \etal{}, it is not specified how the $K$ initial users should be infected, i.e., in our language, how they should be convinced to tweet the message.
In this paper, we focus on a specific user $u_0$ that is trying to maximize his influence and cannot reach other Twitter users through some external communication network.
Then, $u_0$ can only carefully select a given set of users to follow and hope that these users will follow $u_0$ back and will eventually retweet $u_0$'s tweets.
The strategic choice of $u_0$ is then the selection of his set of followings in order to maximize his influence.
We consider for the moment that $u_0$ makes this choice once and for all at his registration.
We observe that Twitter puts a cap to the maximum number of initial followings that is $K=2000$ (this limit increases linearly with the number of followers).
More formally, let $B$ denote the set of $u_0$'s followings and let $\varphi(B)$ be the set of nodes that retweet a tweet originally emitted from $u_0$.
We can write $\varphi(B)=\sum_{v \in V} X_v$, where $X_v$ is a Bernoulli random variable, that is equal to $1$ iff node $v$ is active at the end of the cascade.
Our problem can be formally stated as follows:
\begin{equation}\label{minim1}
\begin{aligned}
& \underset{B}{\text{Argmax}}
& & \E{|\varphi(B)|}\\
& \text{subject to}
& & |B| \le K.
\end{aligned}
\end{equation}
In the same spirit of Kempe \etal{} \cite{Kempe03,Kempe05} we assume to know 1)~the probability $r_u$ that a given user $u$ reciprocates $u_0$ if $u_0$ follows $u$ and 2)~the probability $p_v(u,S)$ that node $v$ reads and retweets the message tweeted or retweeted by $u$, given that the nodes in $S$ have already tweeted or retweeted it.
The knowledge of $p_v(u,S)$ is also required for $u=u_0$ or $u_0 \in S$.

\begin{figure}[!t]
\centering
\subfloat[reciprocation]{\includegraphics[width=1.6in]{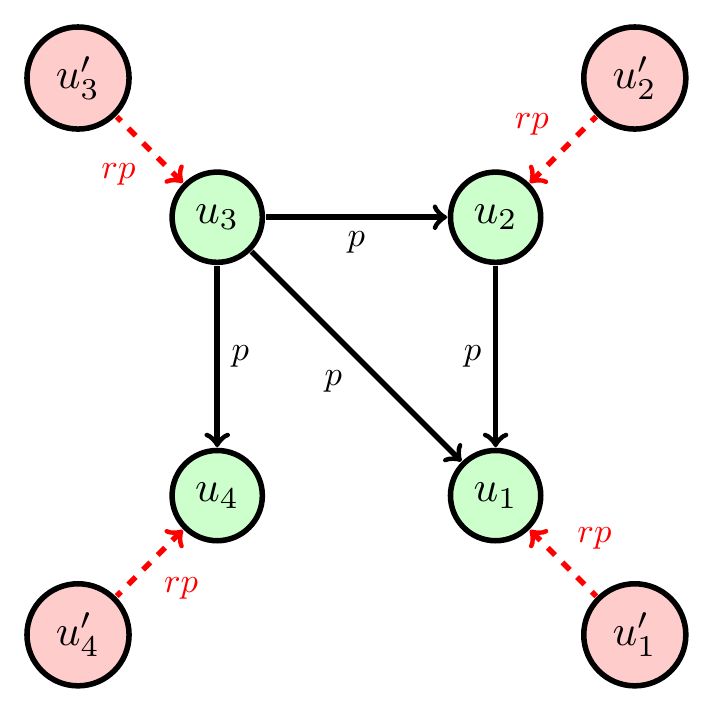}}
\hspace{0.25cm}
\subfloat[readers]{\includegraphics[width=1.6in]{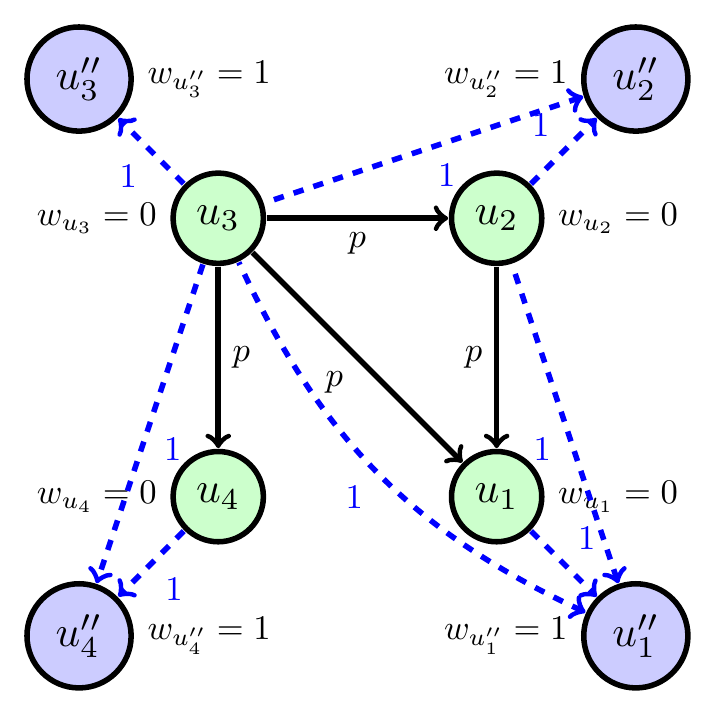}}
\caption{Graph transformations.
The original nodes are green.
The added nodes are red/blue, added arcs are dashed.
We have specified the new success probability functions for the simple case when $p_v(.)$ and $r_v$ are constant and respectively equal to $p$ and $r$.}
\label{fig:transform}
\end{figure}

The greedy algorithm for the follow back problem~\eqref{minim1} corresponds to the following behavior: user $u_0$ selects $K$ followings one after the other, maximizing at each step the marginal increment of the function $\E{|\varphi(.)|}$.
Our first theoretical result is the following.
\begin{prop}
\label{prop1}
The greedy algorithm is a $(1-1/e)$ approximation algorithm for follow back problem~\eqref{minim1}.
\end{prop}
\begin{IEEEproof}
Let $(V,E)$ be the social network's graph without node $u_0$.
Consider a new graph $(\hat V,\hat E)$, where for each node $u$ in $V$ we add a new node $u'$ and a link oriented from $u'$ to $u$.
Let $V'$ be the set of these newly added nodes and $h:V \to V'$ the function such that $h(u)=u'$.
On this graph we define success probability functions as follows: $\hat p_v(t,S)=p_v(t,S)$ for $t \neq v'$ and $v'\notin S$, $\hat p_v(v',S)=r_v p_v(u_0,S)$ and $\hat p_v(t,S)=r_v p_v(t,(S\cup\{u_0\})-\{v'\})$ if $v'=h(v) \in S$.
Fig.~\ref{fig:transform}~(a) illustrates the graph transformation when $r_v=r$ and $p_v(u,S)=p$.
We consider now the order independent cascade model on the graph $(\hat V,\hat E)$.
$u_0$'s choice of the set of his followings $B \subseteq V$ corresponds to the choice of the set $A=h(B)$ of initial active nodes in $V'$.
Moreover, the probabilities have been defined in such a way that it is possible to couple the two processes so that $\phi(h(B))=\varphi(B)+K$, where adding $K$ corresponds to the fact that the initial set of active nodes is counted by $\phi(.)$ ($A\subseteq \phi(A)$).
It follows that $\sigma(h(B))=\E{\varphi(B)}+K$ and the problem~\eqref{minim1} is equivalent to solve the influence maximization problem on $(\hat V,\hat E)$ with the additional constraint that the nodes can only be selected in $V'$.
This does not change the property of the function $\sigma(.)$, that is non negative, monotone, and sub-modular, then the results in~\cite{Kempe03,Kempe05} still hold.
In particular the greedy algorithm is a $(1-1/e)$ approximation algorithm for the influence maximization problem defined on $(\hat V, \hat E)$ and then for the problem~\eqref{minim1}.
\footnote{ Also the results for the case when the greedy algorithm selects at each step a $(1-\epsilon)$ approximation of the best node can be extended to    our case, but for the sake of conciseness, we only refer to the simpler case.}
\end{IEEEproof}

\paragraph{\textbf{Dynamic problem}}
We now consider a variation of the follow back problem~\eqref{minim1} where node $u_0$ is not required to select all the $K$ followings at once, but $u_0$ can apply more complex dynamic strategies.
For example, node $u_0$ can stop following nodes that do not reciprocate by a given time $T$ and start following new users.
In this way, $u_0$ can follow during a given time window more than $K$ users (but at most $K$ at the same time) and reach in general a larger number of followers (the number can approach $K$ if there are at least $K$ nodes in the network willing to reciprocate $u_0$).
This improvement in comparison to the original problem is obtained at the price of a longer time required to select the best followings.
The best possible result achievable by $u_0$ is obtained if we assume $u_0$ to know \emph{a priori} which nodes would reciprocate.
For each node $v$, let $R_v$ be the Bernoulli random variable indicating if node $v$ reciprocates node $u_0$ by time $T$ after $u_0$ starts following $v$.
Clearly it holds $r_v=\E{R_v}$.
We introduce then the following ideal optimization problem:
\begin{equation}\label{minim2}
\begin{aligned}
& \underset{B}{\text{Argmax}}
& & \E{|\varphi(B)|}\\
& \text{subject to}
& & |B| \le K \text{ and } R_v=1 \;\; \forall v \in B.
\end{aligned}
\end{equation}

The greedy algorithm for this problem at each step selects the reciprocating node that maximizes the marginal improvement of $\varphi(B)$.
\begin{prop}
\label{prop2}
The greedy algorithm is a $(1-1/e)$ approximation algorithm for the dynamic problem~\eqref{minim2}.
\end{prop}
\begin{IEEEproof}
The proof is analogous to that of Proposition~\ref{prop1}.
In this case an additional node $u'$ is added only for each reciprocating node $u$, i.e., for each $u \in V$ such that $R_u=1$, and the probabilities can be updated as follows: $\hat p_v(t,S)=p_v(t,S)$ for $t \neq v'$ and $v'\notin S$ , $\hat p_v(v',S)=p_v(u_0,S)$ and $\hat p_v(t,(S \cup \{u_0\})-\{v'\})=p_v(t,S)$ if $v' \in S$, where only the nodes $v'$ such that $R_v=1$ need to be considered.
\end{IEEEproof}

While the problem~\eqref{minim2} requires to know a priori which users are willing to reciprocate $u_0$, the greedy algorithm can be implemented online without such knowledge.
This practical greedy algorithm operates in steps, where each step has a duration at most equal to $T$ time units.
At each step the user follows the node $v$ that brings the largest marginal increase in comparison to the already selected nodes assuming that $v$ reciprocates.
If node $v$ reciprocates by time $T$, node $u_0$ maintains user $v$ in his list of followings, otherwise $u_0$ removes $v$.
The algorithm stops when $K$ users reciprocate or when there are no more users to select in the network.
It is easy to check that the practical greedy algorithm selects exactly the same users that the greedy algorithm with a priori knowledge of the reciprocating nodes would, but it requires in general a longer time to execute. 
The reasoning above leads us to conclude that:

\begin{prop}
\label{prop3}
The greedy algorithm for the problem formalized in Eq.~\ref{minim2} can be implemented without a priori knowledge of which users reciprocate, and its expected number of retweets is at least $(1-1/e)$ of the value obtained by any online algorithm where each node can be selected at most once and reciprocation delays of at most $T$ time units are tolerated.
\end{prop}

\paragraph{\textbf{Reader problem}}
Now, we define the influence of a user $u$ as the average number of users who read $u$'s tweets, because they follow $u$ or because they follow someone who has retweeted one of $u$'s tweets.
This problem can be mapped to a variant of the previous case (where we consider the number of retweets) introducing opportune nodes' weights.
We need to change the original graph $(V,E)$ as follows.
For each user $u \in V$, we introduce a new node $u''$ and the directed arcs $(u,u'')$ and $(v,u'')$, for each node $v$ such that $(v,u) \in E$ (see Fig.~\ref{fig:transform} (b)).
We denote $V''$ and $E''$ respectively the set of new nodes and arcs and $(\tilde V, \tilde E)$ the new graph.
By doubling each node, we can separately account for the two roles of a user as a retweeter and as a reader.
Going back to the cascade model terminology, at a given time step if node $u$ is active, the corresponding user has retweeted the tweet, and if node $u''$ is active the corresponding user has read the tweet.
In order to correctly model the process, we introduce activation success probabilities as follows: $\tilde p_v(t,S)=p_v(t,S)$ for $v \in V$ and $\tilde p_{v''}(t,S)=1$ for $v'' \in V''$.
We also introduce nodes' weights $w_{v}=0$ for $v \in V$ $w_{v''}=1$ for $v''\in V''$.
Let $X_v$ be the Bernoulli random variable that indicates if node $v$ is active when the cascade terminates.
The number of users that see the tweet is given by $\psi(B)=\sum_{v \in \tilde V} w_v X_v$ where $B\subseteq V$ is the set of followings selected by node $u_0$.
Two different problems can then be defined depending if the set $B$ has to be selected at the begin or can be changed dynamically, similarly to what is done
above.
The only difference is the fact that the weighted objective function $\E{\psi(B)}$ is considered instead of the unweighted one $\E{\varphi(B)}$. 
Obviously the function $\E{\psi(B)}$ is non-negative and non-decreasing, we can also prove the following result.
\begin{prop}
\label{p:submodularity}
The function $\E{\psi(B)}$ is submodular.
\end{prop}
\begin{IEEEproof}
We adapt some results of Kempe \etal{}~\cite{Kempe05} relative to the size of the different sets to the case where we consider a weighted sum of the set elements.
We need to prove that $\psi(B_1 \cup \{z\})-\psi(B_1) \ge \psi(B_2 \cup \{z\})-\psi(B_2)$ for any $z$ whenever $B_1\subseteq B_2$.
Let $C=\phi(B)$ be the (random)  set of nodes active at the end of the cascade starting from the nodes in $B$.
Imagine now to start a new cascade process on the graph activating node $z$, but taking into account the fact that all the nodes in $C$ have already tried to infect their neighbors.
This new cascade is called the \emph{residual cascade process} and has success probabilities $p_v^{(C)}(u,S)\triangleq  p_v(u,S \cup C)$.
We denote this new stochastic process as $\mathcal S_C(z)$ and the additional nodes in $V \setminus C$ made active by it as $\phi_C(z)$.
Kempe \etal{} \cite[Theorem~3]{Kempe05} proved that $\phi_C(z)$ is distributed as $\phi(B \cup \{z\})-\phi(B)$.
Then it holds:
\begin{equation}
\label{e:submodular_first}
\E{\psi(B \cup \{z\})-\psi(B)}=\E{\sum_{v\in \phi_{\phi(B)}(z)} w_v}
\end{equation}
Consider $C_1 \subseteq C_2$, and the corresponding residual processes $\mathcal S_{C_1}(z)$ and $\mathcal S_{C_2}(z)$.
If we couple the equivalent general threshold models by selecting the same threshold at each node, Kempe \etal{}~\cite[Lemma~3]{Kempe05} showed that pathwise $\phi_{C_1}(z)\supseteq \phi_{C_2}(z)$. It follows that $\sum_{v \in \phi_{C_1}(z)} w_v \ge \sum_{v \in \phi_{C_2}(z)} w_v$ and then
\begin{equation}
\label{e:submodular_second}
\E{\sum_{v \in \phi_{C_1}(z)} w_v} \ge \E{\sum_{v \in \phi_{C_2}(z)} w_v} \text{ whenever }C_1 \subseteq C_2.
\end{equation} 	
Let us now consider two cascade processes whose initial activation sets are respectively $B_1$ and $B_2$ with $B_1\subseteq B_2$, if we couple them as above, we can similarly show that $\phi(B_1) \subseteq \phi(B_2)$, then
\begin{equation}
\label{e:submodular_third}
P(\phi(B_1)=C_1,\phi(B_2)=C_2)=0\textrm{  whenever }C_1 \not \subseteq C_2.
\end{equation}
We can now wrap-up our intermediate results. Let $B_1\subseteq B_2$, then
\begin{eqnarray*}
\lefteqn{ \E{\psi(B_1 \cup \{z\})-\psi(B_1)}=\textrm{E}\!\left[\sum_{v\in \phi_{\phi(B_1)}(z)} \!\!\! w_v\right]}\\
&=&\sum_{C_1}\textrm{E}\!\left[\sum_{v\in \phi_{C_1}(z)} \!\!\! w_v\right] P(\phi(B_1)=C_1)\\
&=&\sum_{C_1}\sum_{C_2 \supseteq C_1}\textrm{E}\!\left[\sum_{v\in \phi_{C_1}(z)}\!\!\! w_v\right] P(\phi(B_1)\!=\!C_1,\phi(B_2)\!=\!C_2)\\
&\ge&\sum_{C_1}\sum_{C_2 \supseteq C_1}\textrm{E}\!\left[\sum_{v\in \phi_{C_2}(z)}\!\!\! w_v\right] P(\phi(B_1)\!=\!C_1,\phi(B_2)\!=\!C_2)\\
&=&\sum_{C_2} \textrm{E}\!\left[\sum_{v\in \phi_{C_2}(z)}\!\!\! w_v\right] P(\phi(B_2)=C_2)\\
&=&\E{\psi(B_2 \cup \{z\})-\psi(B_2)},
\end{eqnarray*}
where we have used   Eqs.~\eqref{e:submodular_first},~\eqref{e:submodular_second}~and~\eqref{e:submodular_third}.
\end{IEEEproof}
From the general results for non-negative, non-decreasing submodular functions, it follows that the greedy algorithm that selects orderly the  $u_0$'s followings that incrementally maximize $\E{\psi(.)}$ guarantees a $(1-1/e)$ approximation ratio.
The same result holds in the dynamic case.
Due to lack of space we do not define formally the two problems and the corresponding propositions, but we summarize our conclusions as follows.
\begin{prop}
\label{p:greedy3}
The greedy algorithms for the static and dynamic versions of our problem reach a $(1-1/e)$ approximation ratio also when the objective function is $\E{\psi(.)}$, the expected number of users who see the tweet.
\end{prop}

\section{Simulations on Twitter}
\label{s:experiments}
We have shown in the previous section that for all three specificities of OSNs, the greedy algorithm is a good approximation of the optimal algorithm.
As the optimal algorithm is NP-hard, it is a major improvement.
Unfortunately, the greedy algorithm requires to know the topology of a social graph as well as all the functions $p_v(u,S)$ for every node $v$, a requirement that is not feasible for the social graph of the size of Twitter.
Moreover, greedy algorithms are computationally expensive, because of the inherent cost of evaluating the expected size of $\phi(A)$, that can only be estimated by Monte Carlo simulations of the cascade process on the social graph.

In this section, we show using simulations on the full Twitter social graph that simple and practically feasible strategies perform close to the greedy algorithm. We describe the methodology we used to perform our simulations in Section~\ref{sec:exp_methodology} and we discuss the results in Section~\ref{sec:exp_results}.

\subsection{Methodology}
\label{sec:exp_methodology}
For our simulations, we considered the simple case when $p_v(u,S) = p$ is constant and evaluated different selection strategies on the complete social graph of Twitter as crawled in July 2012~\cite{Gabielkov12}.
We had to solve two main problems. 

The first problem to solve with the simulations is to manage the large size of the Twitter social graph.
Indeed, this graph consists of 505 million nodes and 23 billion arcs and requires roughly 417GB of storage in the form of edgelist.
A naive implementation of the simulation would require to load the graph into memory and then use Monte Carlo simulations of the retweet process in order to estimate the objective functions with high accuracy.
The followers of the initial node $u_0$ would retweet with probability $1$, then their followers will retweet with probability $p$ and so on until no new node is retweeting.
To reduce the memory required to make this computation, we introduce the concept of pruned graphs.
A \emph{pruned} graph is obtained from the original one by sampling each edge with probability $p$ (and with probability $1-p$ the edge is removed from the graph).
Computing the set of reachable nodes from $u_0$ on a pruned graph is equivalent to counting the number of retweeting nodes in a specific sample of the retweet random process, but memory requirement is reduced by a factor $p$ (usually $p<<1$) at the expense of storage increase, because we need to work on multiple pruned graphs (see discussion below) in order to reach the required accuracy.

However, for large values of $p$ the pruned graphs are still large, so to reduce further the memory requirement, we use a two step compression.
First we compute the Strongly Connected Components (SCCs) of the graph.
Second, we construct a Directed Acyclic Graph (DAG) by abstracting each SCC as a single node and replacing multiple arcs between the nodes with a single arc.
Provided that $p$ is quite large, we will observe big SCCs in the pruned graph, thus we can achieve a big reduction in size using our approach.
Then we compute the reachability on the obtained DAG and deduce the reachability of the original pruned graph by taking into account the number of nodes in each SCC and the fact that the nodes belonging to the same SCC have the same reachability.
This approach decreases the computation time, as well as memory and storage requirements (because of the more compact DAG representation). 
Both these expedients were proposed by Chen \etal{} \cite{Chen09}, but we have been able to make computations on a graph 4 orders of magnitude larger.

\switch{}{
\begin{figure}[!t]
\centering
\includegraphics[width=2.5in]{figures/models_samples_vs_retweet_p}
\caption{Models.}
\label{models_samples_vs_retweet_p}
\end{figure}
}

The second problem to solve is to determine how many simulations we should run (that is, how many pruned graph we should compute) in order to achieve a given precision for the estimates of the expected number of retweets.
To this purpose, we have used two different models that we can only describe shortly here because of space constraints (details are available in a technical report~\cite{Ye13}).
The first model approximates the cascade process with a branching process, where the probability to have a follower with $k$ followers is $k q_k/\av{k}$, where $q_k$ is the distribution of the number of followers in the graph and $\av{k}$ is the average number of followers (see for example \cite[Chapter~8]{Barrat2008} for a justification of such an expression).
This model requires that different active nodes have different followers and is good only for small values of $p$.
In particular, it may be accurate only when the branching process dies out with probability $1$, because otherwise the model predicts that the expected number of active nodes is infinite while this number is obviously limited by the total number $N$($=|V|$) of nodes in the graph.
The branching process extinguishes with probability $1$ if $p \sum_k k q_k/\av{k} <1$, i.e.,~$p < 2 \times 10^{-4}$ in the considered Twitter graph.
The second model addresses the case for $p>2 \times 10^{-4}$.
In this case the branching process theory predicts that the process can still extinguish with a probability $p_{ext}$ that is a decreasing function of $p$.
The intuition behind the second model is to couple the branching process and the actual cascade model and assume that the cascade will reach almost all the $N$ nodes when the branching process does not extinguish and a negligible number of nodes when it does.
In particular, given that we are interested in providing upper-bounds for the variability of the process, we simply consider that the number of active nodes is a random variable that is equal to $N$ with probability $1-p_{ext}$ and to $0$ with probability $p_{ext}$.
Some further refinements of the model lead to the conclusion that the number of samples needed to achieve a reasonable prevision is below $100$ for all the values of $p$ we considered, i.e.,~$p=10^{-4}, 10^{-3}, 10^{-2}, 10^{-1}$. This result is quite surprising, given the high variability of the degree distribution $\{q_k\}$ (that is power law) and the even higher variability of the skewed distribution $\{k q_k /\av{k}\}$. 

With the approach described above we have been able to perform our simulations on the \emph{real} Twitter social graph with hundreds of millions of nodes and tens of billions of arcs.

\subsection{Results}
\label{sec:exp_results}

\begin{figure*}[!t]
\centering
\subfloat[$p = 0.0001$]{\includegraphics[width=1.65in]{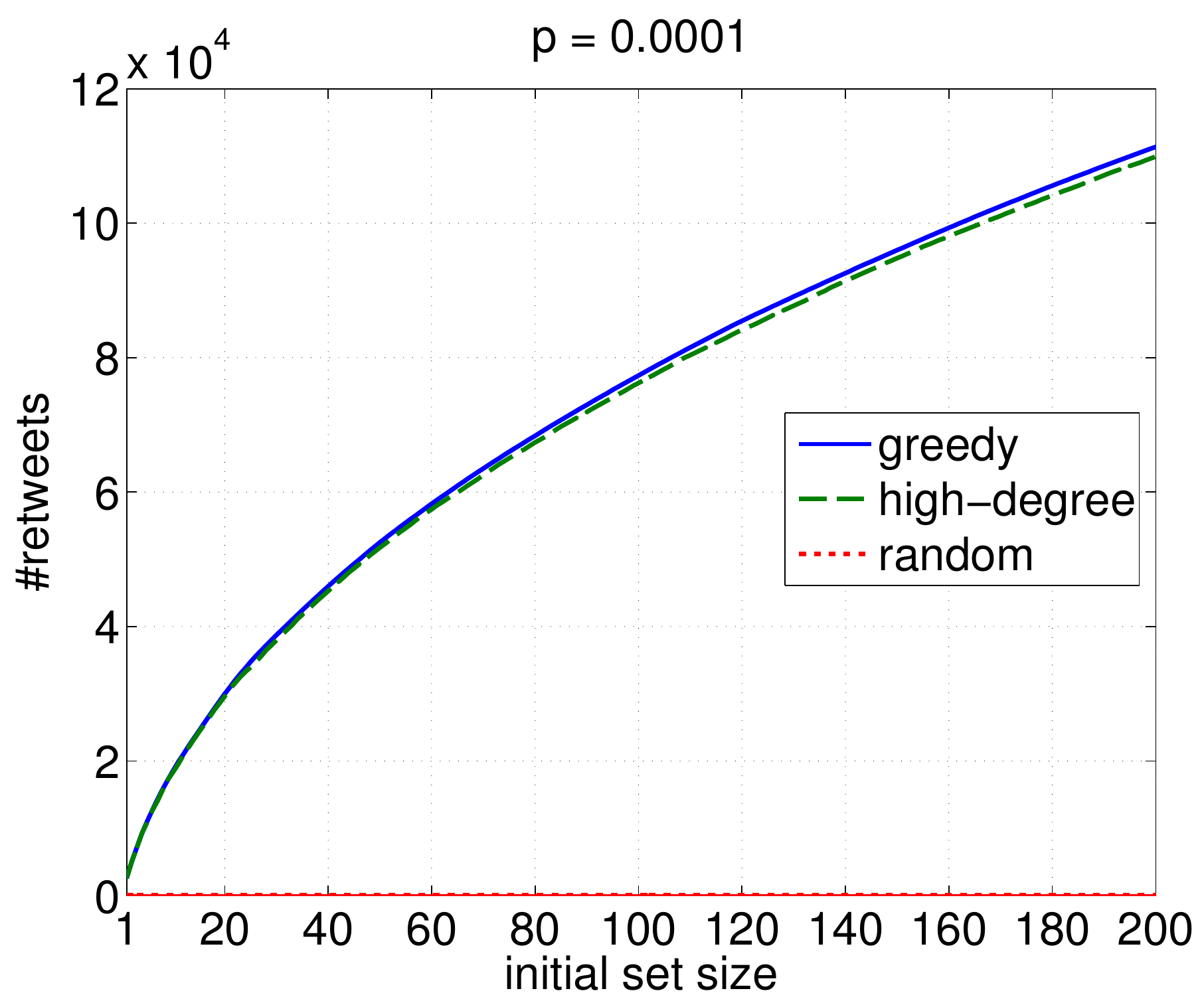}}
\hspace{0.25cm}
\subfloat[$p = 0.001$]{\includegraphics[width=1.70in]{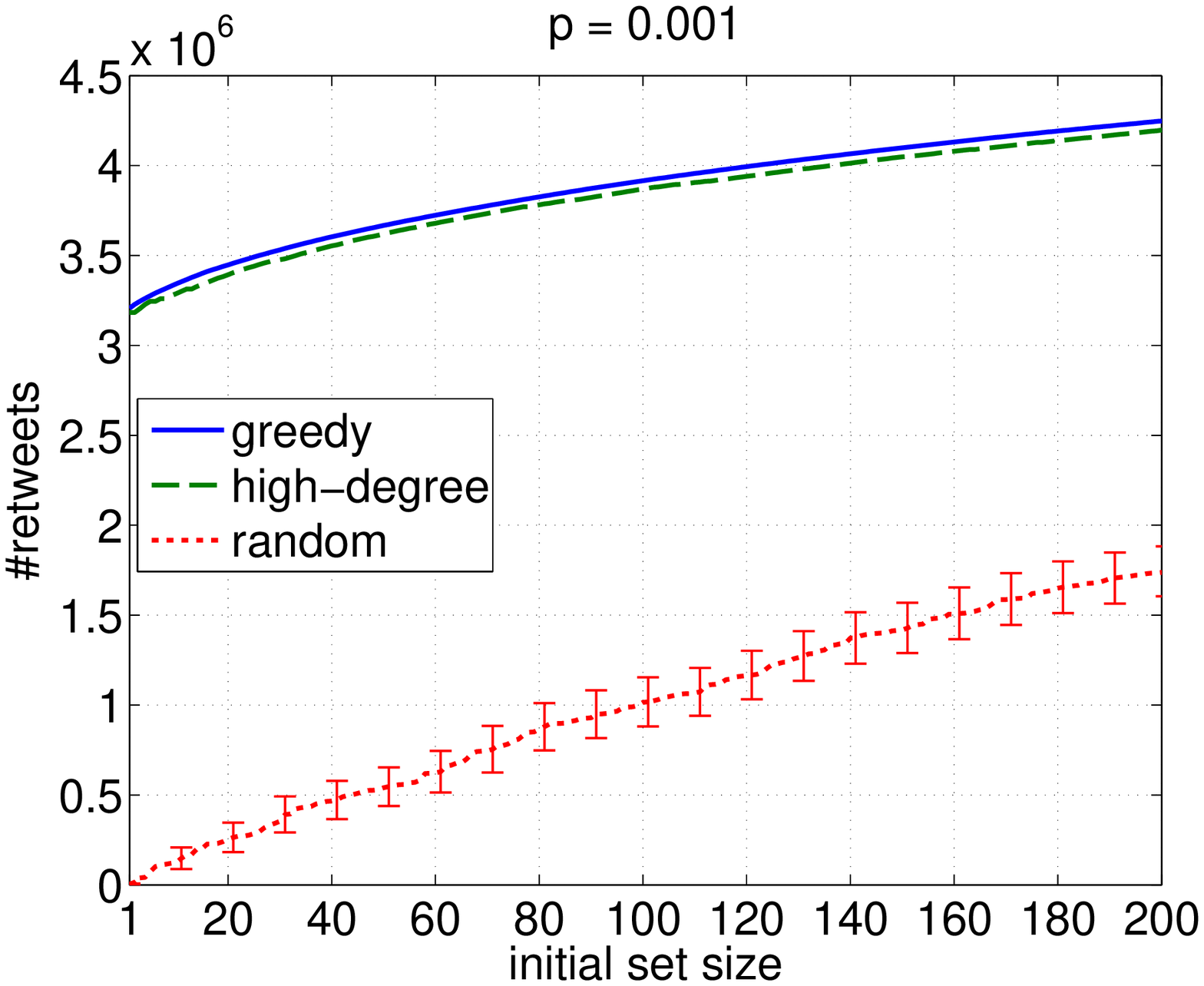}}
\hspace{0.25cm}
\subfloat[$p = 0.01$]{\includegraphics[width=1.65in]{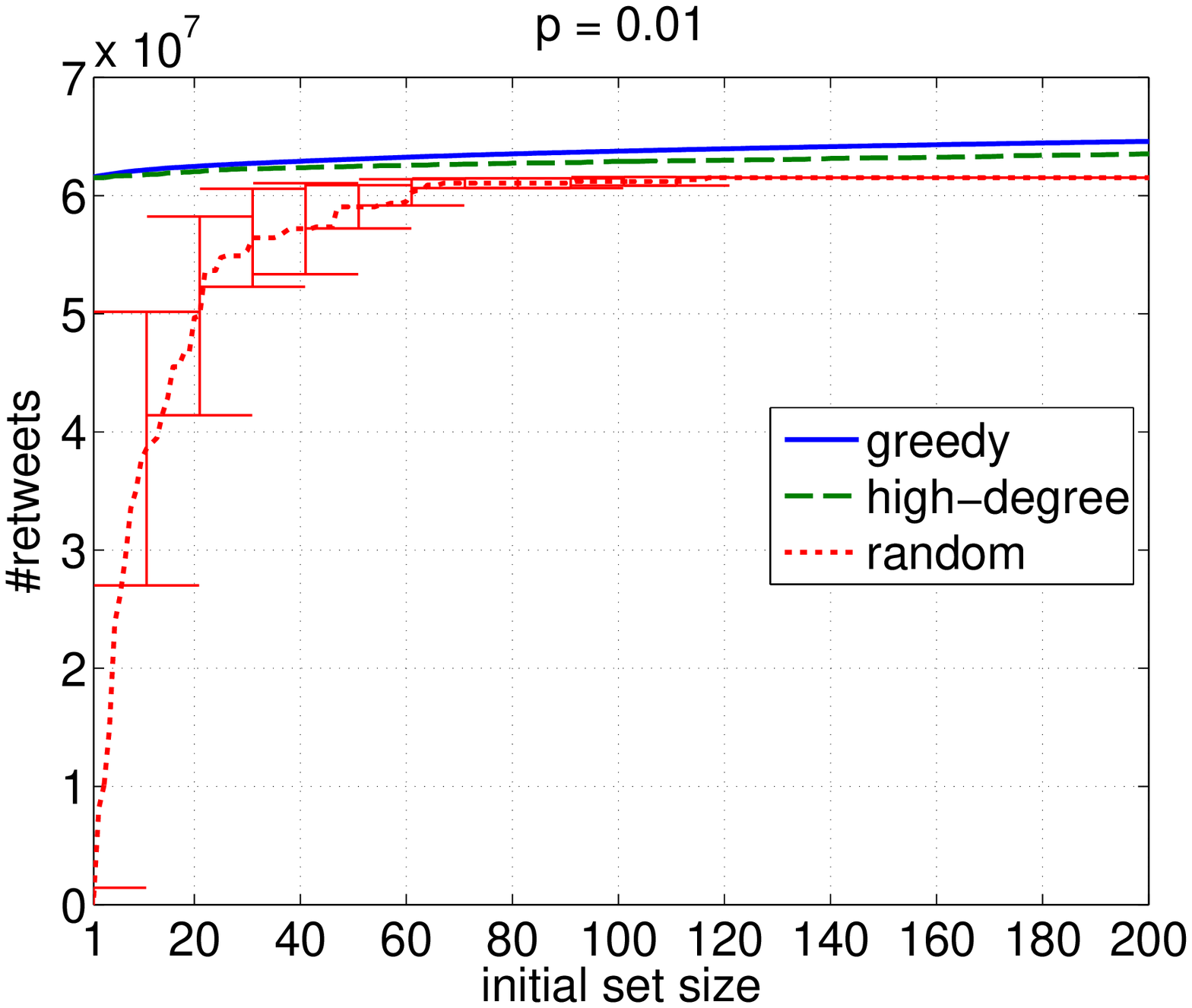}}
\hspace{0.25cm}
\subfloat[$p = 0.1$]{\includegraphics[width=1.70in]{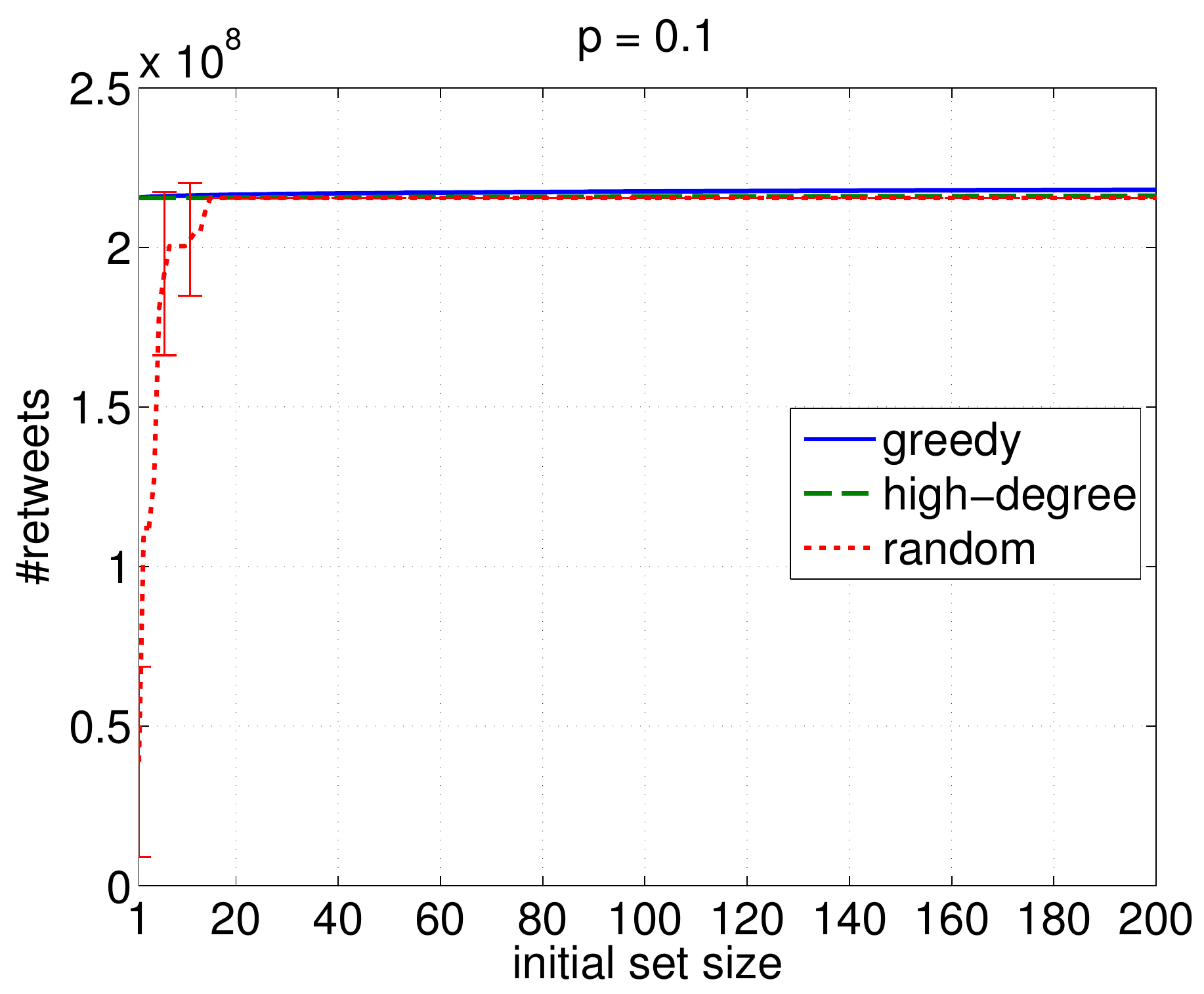}}
\caption{Performance of greedy, high-degree, and random strategy on the   Twitter social graph for different values of retweet probabilities $p$.
The reciprocation probability is $r = 1$. Error bars show the $95\%$ confidence intervals.}
\label{fig:4figures_retweets_vs_initial_set_size}
\end{figure*}

In this section, we compare using simulations three selection strategies for the user $u_0$ to select his followers:
i) the {\bf greedy } strategy described above;
ii) the {\bf high-degree} strategy which consists in picking the nodes orderly according to their number of followers (from the largest number to the smallest);
iii) the {\bf random} strategy where followers are selected uniformly at random from the whole set of users.

We have to evaluate the impact of two main parameters: the retweet probability $p$ and the reciprocation probability $r$.
We start by considering the impact of $p$ on the three selection strategies by assuming $r=1$, i.e., $u_0$ is followed by every node it follows.
While unrealistic, this assumption is an easy way to assess the impact of $p$ only.
Then in a second set of simulations, we relax this assumption and define $r$ as a function of the number of followings and followers of each node.

In our first set of simulations, we consider $r=1$ and that $u_0$'s followers retweet $u_0$'s tweets with probability $1$, while all the other users retweet with probability $p \in \{10^{-4}, 10^{-3}, 10^{-2}, 10^{-1}\}$. 
Fig.~\ref{fig:4figures_retweets_vs_initial_set_size} shows the expected number of retweets versus the initial number $K$ of followers $u_0$ can choose.
The average number of followers in the original graph is $\av{k}\approx45$, but the effective density, as defined by Habiba \etal{} \cite{Habiba11}, is $\av{k} p \approx 4* 10^{-3}$ for $p=10^{-4}$ (in Fig.~\ref{fig:4figures_retweets_vs_initial_set_size}~(a)).
Then, in this case, most of the nodes are not retweeted by any follower, but due to the skewness of the distribution $\{q_k\}$ (the number of followers can be as high as 24,635,412), there are some hubs in the social graph that have an expected number of retweeters significantly larger than $0$.
The cascade processes from different followers of $u_0$ do not overlap much (each pruned graph is almost a forest of small-depth trees with a multitude of singletons), so that the high-degree strategy performs almost as well as the greedy strategy.
Due to this structure, the expected number of retweeters significantly increases as the number of $u_0$'s followers keeps increasing.
The random strategy performs poorly, because with high probability the $200$ selected followers will be singletons.
When $p=10^{-3}$ (Fig.~\ref{fig:4figures_retweets_vs_initial_set_size}~(b)), the cascade originated from the node with the largest degree is already able to reach about $3*10^{6}$ users (roughly $1\%$ of the whole social graph) and both the greedy and high-degree strategy select this node first.
The other followers selected from $u_0$ using these two strategies lead to a minor improvement: even adding $199$ more followers, the expected number of retweeters increases by only $33\%$.
The random strategy starts paying off because there are much less singletons in the pruned graph. Further increasing $p$ to $10^{-2}$, the contribution of the first follower is even larger and the contribution of the others even more marginal, as it is shown in the plot in Fig.~\ref{fig:4figures_retweets_vs_initial_set_size}~(c).
In fact, a non-negligible SCC appears in most of the pruned graphs and a careful choice of the first follower allows $u_0$ to have roughly one tenth of the nodes retweeting his tweets (this follower is not necessarily in the largest SCC, but he can reach it).
The other $199$ followers provide roughly $5\%$ more retweeters.
We observe that the effective density is about $0.4$, then more than half of the nodes have $0$ out-degree/in-degree in the pruned graphs.
The greedy strategy and the high-degree one lead to a difference in the number of retweets lower than $2.5\%$.
Interestingly, the probability to randomly pick a node in the largest SCC is now quite high, so the random strategy performs close to the other two strategies, but with a higher variability as shown by the large confidence intervals in the figure.
Moreover, the figure shows how the good follower is very likely to be selected among the first $10$-$20$ nodes.
The same reasoning allows to explain also the curves in Fig.~\ref{fig:4figures_retweets_vs_initial_set_size}~(d) for $p=0.1$.
In this case, greedy and high-degree are almost indistinguishable and random has almost the same performance for $K\ge20$.

\begin{figure}[!t]
\centering
\subfloat[Reciprocation probability]{\includegraphics[width=1.7in]{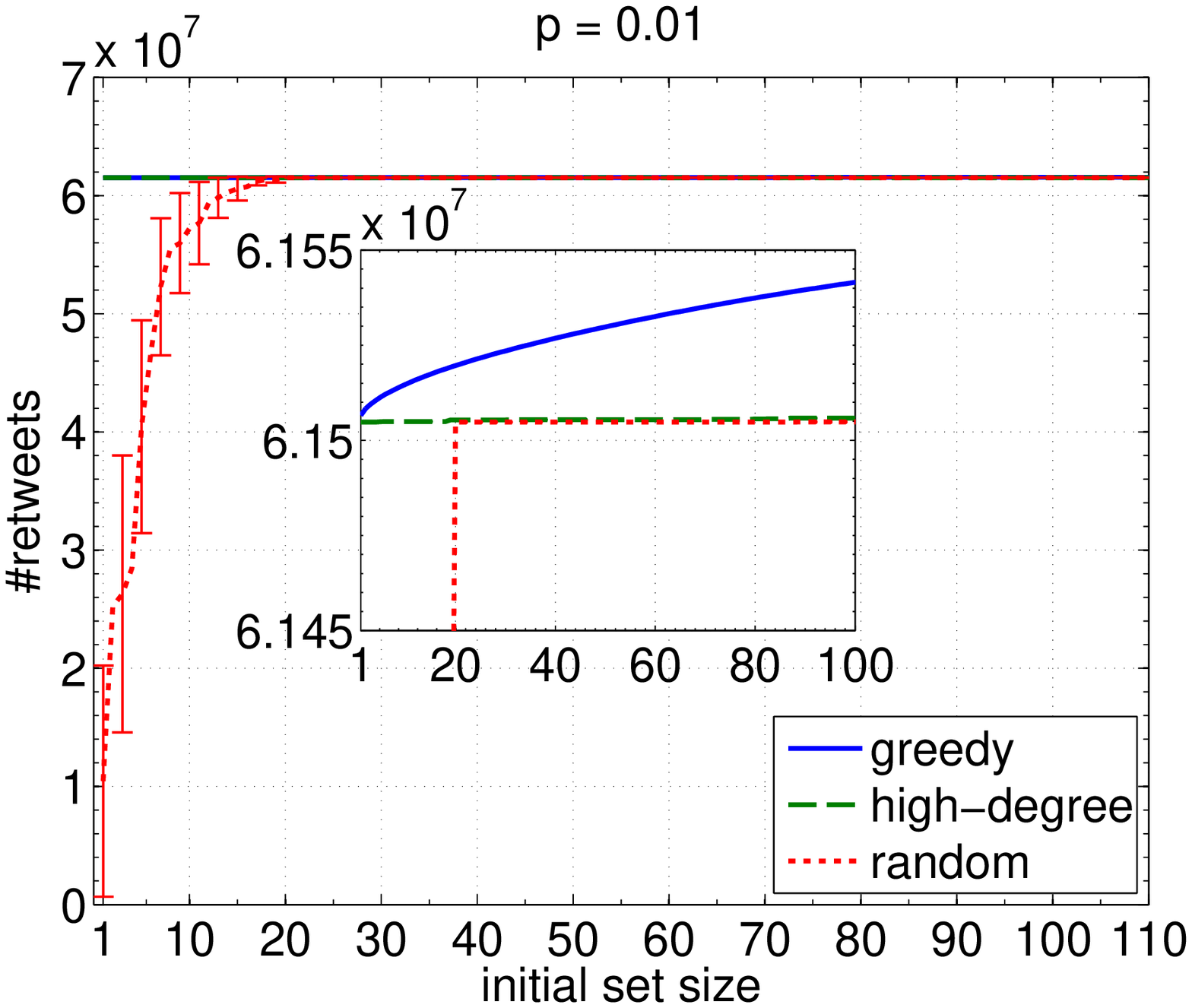}}
\subfloat[Number of readers]{\includegraphics[width=1.7in]{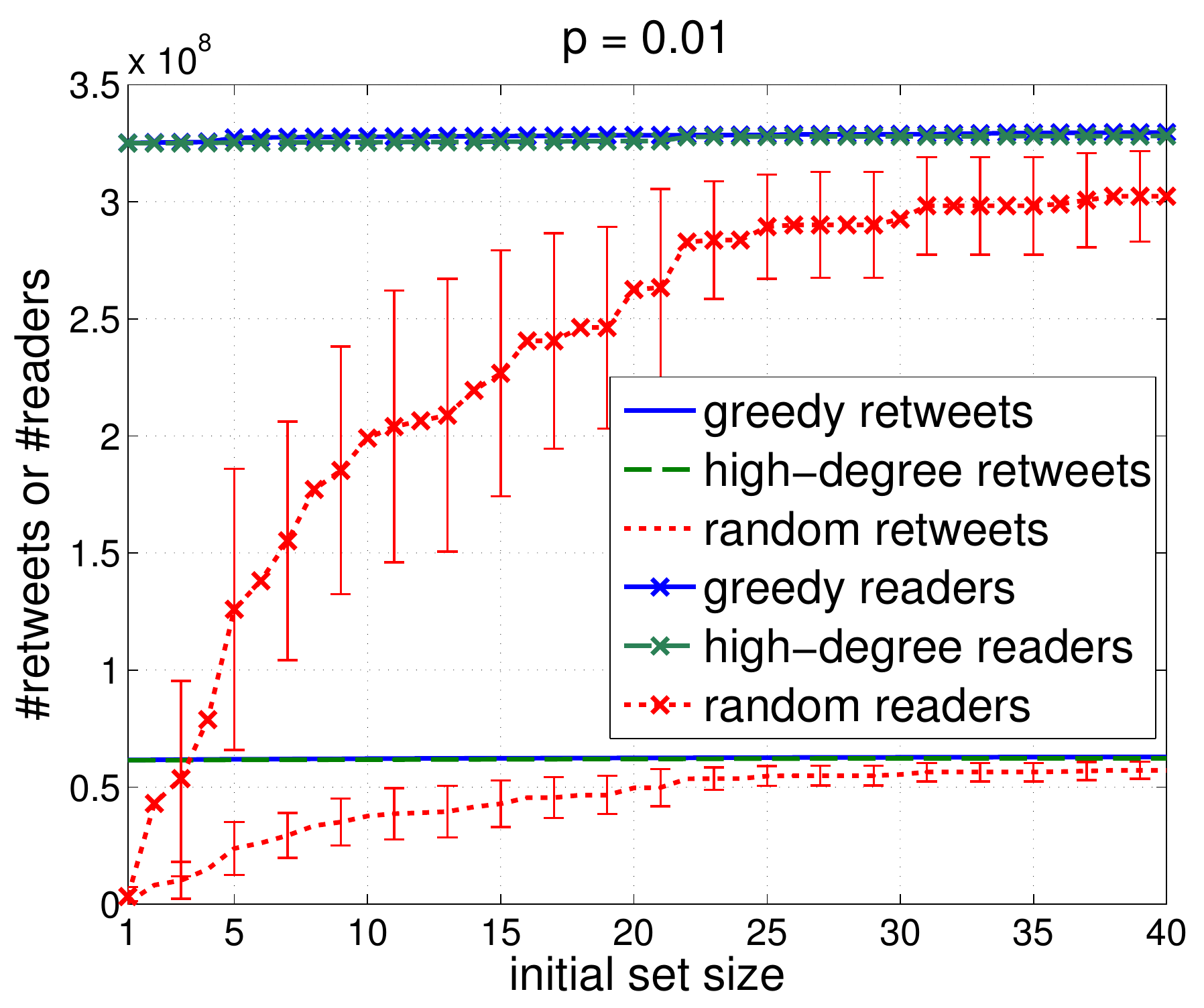}}
\caption{Extensions of the simulations by taking into account the probability $r$ that users will follow back (a), and by looking at the number of users who received the tweet instead of the number of users who retweeted it (b).}
\label{fig:extensions}
\end{figure}

In our second set of simulations, we consider $r_v = \text{min}\{\frac{\text{\#followings}_v}{\text{\#followers}_v+100},1\}$, where $\#\text{followings}_v$ and $\#\text{followers}_v$ are respectively the number of followings and followers for user $v$.
The rationale behind is that a user with a lot of followers and a few followings is not likely to reciprocate $u_0$.
We do not claim that this formula has any particular value, apart from allowing us to simply test the effect of heterogeneous reciprocation probabilities.
In this case the high-degree strategy selects the nodes according to their \textit{effective degree} $r_i d_i$. 
Surprisingly, the results for $p=0.01$ are qualitatively unchanged as shown in Fig.~\ref{fig:extensions} (a).
So the reciprocation probability does not seem to significantly impact the respective performance of the three considered strategies.
We have also compared the different algorithms in terms of the expected number of users who can read the tweet.
We see in Fig.~\ref{fig:extensions} (b) that the number of readers is obviously much bigger than the number of retweeters, but there is no significant difference in the relative performance of the three strategies.

\section{Conclusions}
In this paper we have considered a user of a social network who tries to maximize his influence through a careful networking strategy.
We have shown how greedy algorithms guarantee a good $1-1/e$ approximation ratio, but much simpler strategies like selecting users with the largest number of followers or even selecting random users may practically reach the same performance on real online social networks.

This research is partially supported by Alcatel Lucent Bell Labs in the framework of the ADR Network Science. 
The authors would like to thank Alonso Silva (Alcatel Lucent Bell Labs), Paolo Giaccone (Politecnico di Torino) and Damien Saucez (Inria) for the helpful discussions.

\bibliographystyle{./IEEEtran}
\bibliography{./IEEEabrv,./influence}

\end{document}